# Weighted finite impulse response filter for chromatic dispersion equalization in coherent optical fiber communication systems

ZIYI ZENG,[1] AIYING YANG,[1,*] PENG GUO[1,*], LIHUI FENG[1]

[1] *School of Optoelectronics, Beijing Institute of Technology, Beijing 100081, China*
*\*yangaiying@bit.edu.cn, guopeng0304@bit.edu.cn*

**Abstract:** Time-domain chromatic dispersion (CD) equalization using finite impulse response (FIR) filter is now a common approach for coherent optical fiber communication systems. The complex weights of FIR filter taps are calculated from a truncated impulse response of the CD transfer function, and the modulus of the complex weights is constant. In our work, we take the limited bandwidth of a single channel signal into account and propose weighted FIR filters to improve the performance of CD equalization. A raised cosine FIR filter and a Gaussian FIR filter are investigated in our work. The optimization of raised cosine FIR filter and Gaussian FIR filter are made in terms of the EVM of QPSK, 16QAM and 32QAM coherent detection signal. The results demonstrate that the optimized parameters of the weighted filters are independent of the modulation format, symbol rate and the length of transmission fiber. With the optimized weighted FIR filters, the EVM of CD equalization signal is decreased significantly. The principle of weighted FIR filter can also be extended to other symmetric functions as weighted functions.

## 1. Introduction

Chromatic dispersion (CD) was once considered one of the severe impairments limiting optical fiber communication systems [1]. Fortunately, with the advancement of digital signal processing (DSP) technology, coherent optical receivers exploiting digital filters can effectively equalize CD of an optical fiber communication system [2-11]. Since the fiber CD transfer function is $H(\omega) = \exp\left(-j\frac{D\lambda^2 L}{4\pi c}\omega^2\right)$, the dispersion equalization filter is given by an all-pass filter $H_{eq}(\omega) = \frac{1}{H(\omega)}$, which is not possible to be designed in reality. In order to solve this problem, S. Tsukamoto presented a transversal filter and S. J. Savory presented a finite impulse response (FIR) filter to equalize the CD in time domain for long-haul optical fiber communication systems [12-14]. In Tsukamoto's method, the complex tap weights $c_k$ of a transversal filter is obtained from the inverted Fourier transform of a truncated $H_{eq}(\omega)$, and the truncated window is determined by the bandwidth of a single channel signal [12]. In Savory's method, the complex tap weights $c_k$ of a FIR filter are obtained from the truncated impulse response of $h_{eq}(t) = \sqrt{\frac{jc}{D\lambda^2 L}}\exp\left(-j\frac{\pi c}{D\lambda^2 L}t^2\right)$ [14]. To avoid frequency aliasing, an upper bound on the number of the taps is determined by the accumulated dispersion and sampling interval as $c_k = \sqrt{\frac{jcT_S}{D\lambda^2 L}}\exp\left(-j\frac{\pi c T_S^2}{D\lambda^2 L}k^2\right)$, with $-\left\lfloor\frac{N}{2}\right\rfloor \leq k \leq \left\lfloor\frac{N}{2}\right\rfloor$ and $N = 2\times\left\lfloor\frac{|D|\lambda^2 L}{2cT_S^2}\right\rfloor + 1$. The truncated $H_{eq}(\omega)$ or truncated $h_{eq}(t)$ can be viewed as the rectangular window operation in frequency domain or time domain. In our work, we take the limited bandwidth of a single channel signal into account and propose a weighted FIR filter to avoid frequency aliasing and further improve the performance of CD equalization. The remainder of the paper is organized as the following: we described the principle of a weighted FIR filter in section 2. Then, we investigated two weighted FIR filters, raised cosine type and Gaussian type, in section 3. The critical parameters of two type weighted FIR filters are optimized with VPI simulation in this section. The performance of coherent optical fiber communication systems with weighted filters are demonstrated in section 4. Finally, the conclusions are drawn.

## 2. Principle of a weighted FIR filter

For a weighted FIR filter, the tap weights are given by

$$a_k = \sqrt{\frac{jcT_S}{D\lambda^2 L}}\exp\left(-j\frac{\pi c T_S^2}{D\lambda^2 L}k^2\right)\cdot f(k) \tag{1}$$

where $f(k)$ is sampled from a weighted function $f(t)$. Since the impulse response $h_{eq}(t) = \sqrt{\dfrac{jc}{D\lambda^2 L}} \exp\left(-j\dfrac{\pi c}{D\lambda^2 L} t^2\right)$ is symmetric, $f(t)$ can be a raised cosine function or a Gaussian function.

A raised cosine function $f_{RC}(t)$ is described as

$$f_{RC}(t) = \begin{cases} 1 & 0 \leq |t| < \dfrac{1-\alpha}{2} \cdot T \\ \dfrac{1}{2}\left\{1 + \cos\left[\dfrac{\pi}{\alpha \cdot T}\left(|t| - \dfrac{1-\alpha}{2} \cdot T\right)\right]\right\} & \dfrac{1-\alpha}{2} \cdot T \leq |t| \leq \dfrac{1+\alpha}{2} \cdot T \\ 0 & |t| > \dfrac{1+\alpha}{2} \cdot T \end{cases} \quad (2)$$

where $\alpha$ is the roll-off factor. $T$ is the full width at half maximum (FWHM) of $f_{RC}(t)$. To avoid frequency aliasing, time length of the truncation window of a FIR filter, $T_{window}$, is given by [14, 15]

$$T_{window} = \dfrac{|D|\lambda^2 L}{cT_S}. \quad (3)$$

As in [14, 15], the number of FIR filter taps is $N = 2 \times \left\lfloor \dfrac{|D|\lambda^2 L}{2cT_S^2} \right\rfloor + 1$. To set $t = t_k = 2k \cdot \dfrac{T_{window}}{N}$, with $-\left\lfloor \dfrac{N}{2} \right\rfloor \leq k \leq \left\lfloor \dfrac{N}{2} \right\rfloor$ we can get $f_{RC}(k)$ with Eq. (2). It is not hard to understand that $T$ is related with the truncation window of a FIR filter. To define the window coefficient $X_{RC}$ of a raised cosine function, there is

$$T = X_{RC} \cdot T_{window}. \quad (4)$$

A Gaussian function $f_{GS}(t)$ is given by

$$f_{GS}(t) = \exp\left(-\dfrac{t^2}{2T_0^2}\right) \quad (5)$$

where $T_0$ is the half-width at 1/e-intensity point. The FWHM of $f_{GS}(t)$ is expressed in term of $T_0$ as

$$T = 2(\ln 2)^{1/2} T_0. \quad (6)$$

Similar to raised cosine function, we can get $f_{RC}(k)$ with $t = 2k \cdot \dfrac{T_{window}}{N}$. To define the window coefficient $X_{GS}$ of a Gaussian function, there is

$$T = X_{GS} \cdot T_{window}. \tag{7}$$

To substitute $f_{RC}(k)$ or $f_{GS}(k)$ into Eq. (1), we can get the tap weights of raised cosine weighted FIR filter (RC-FIR filter) or Gaussian weighted FIR filter (GS-FIR filter). Then, the output of the weighted FIR filter is

$$E_{eq}(n) = \sum_{k=0}^{N-1} a_k E(n-k). \tag{8}$$

## 3. Optimization of weighted FIR filter

The tap weights of RC-FIR filter is determined by the roll-off factor $\alpha$ and the FWHM ($T$) of a raised cosine function $f_{RC}(t)$. And the tap weights of GS-FIR filter is determined by $T$ of Gaussian funciton $f_{GS}(t)$. Thus, the RC-FIR filter can be optimized by the roll-off factor $\alpha$ and $T$, i.e. $X_{GS}$, while the GS-FIR filter can be optimized by $T$, i.e. $X_{GS}$. Error vector magnitude (EVM) is normally a critical parameter to evaluate the performance of a coherent communication system [16, 17]. Thus, the optimization of RC-FIR filter, i.e. $\alpha$ and $X_{RC}$ is performed based on the minimum EVM of a CD equalization optical communication system. Fig. 1 is the schematic setup of a coherent optical fiber communication system. The launched optical power of SSMF is 0.0dBm. The chromatic dispersion of SSMF is 16.0ps/(nm.km). The fiber length of each span is 100km. The inline EDFAs are used to compensate the fiber loss of each fiber span. At the coherent receiver, the sample rate of analog-to-digital convertors (ADCs) is twice the symbol rate. The sampled signals are processed by the DSP equalizer.

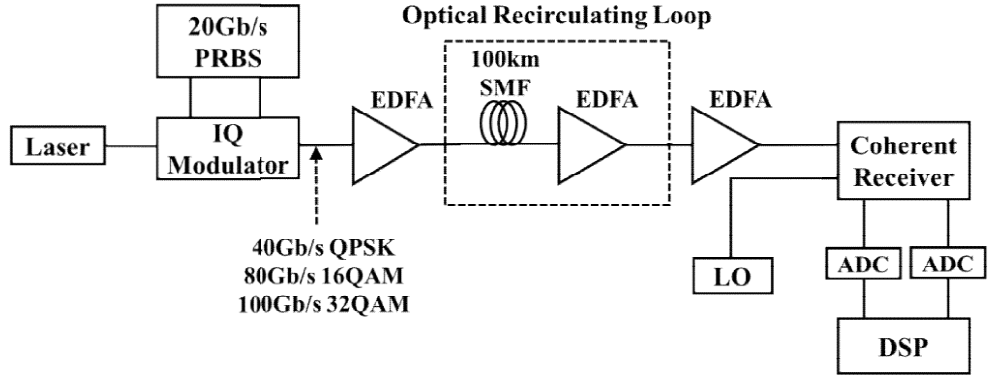

Fig. 1. Schematic setup of a coherent optical transmission system.

The optimization of weighted filter is based on the simulation results with the VPI software. Figs. 2-4 demonstrate the EVM of a QPSK optical fiber transmission system with a RC-FIR filter. The total standard single mode fiber (SSMF) length is 500 km, 1000 km and 1500 km. The symbol rate is 10GBaud, 20GBaud and 30GBaud respectively. As a critical parameter representing the performance of an optical fiber communication system, smaller EVM is preferred. The bar on the right of each figure describes the percentage of EVM. The EVM contour with varying $\alpha$ and $X_{RC}$ results indicates that the EVM of RC-FIR filter equalizing CD system is least when $X_{RC} = 0.7$ and $\alpha = 0.3$, regardless of the fiber length and the symbol rate of signal. One exception is that with $X_{RC} = 0.8$ and $\alpha = 0.3$, the EVM of

10GBaud QPSK 500 km SSMF system is merely slightly larger than that with $X_{RC} = 0.7$ and $\alpha = 0.3$. Therefore, the RC-FIR filter can be optimized with $X_{RC} = 0.7$ and $\alpha = 0.3$.

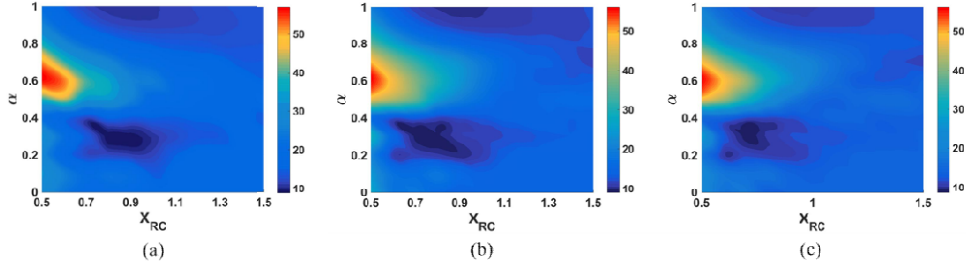

Fig. 2. EVM of a 10GBaud QPSK optical fiber communication system with a RC-FIR filter equalizing CD. (a)500 km SSMF. (b)1000 km SSMF. (c) 1500 km SSMF.

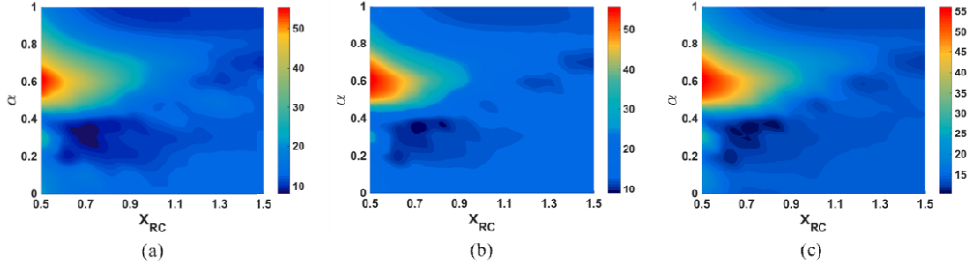

Fig. 3. EVM of a 20GBaud QPSK optical fiber communication system with a RC-FIR filter equalizing CD. (a)500 km SSMF. (b)1000 km SSMF. (c) 1500 km SSMF.

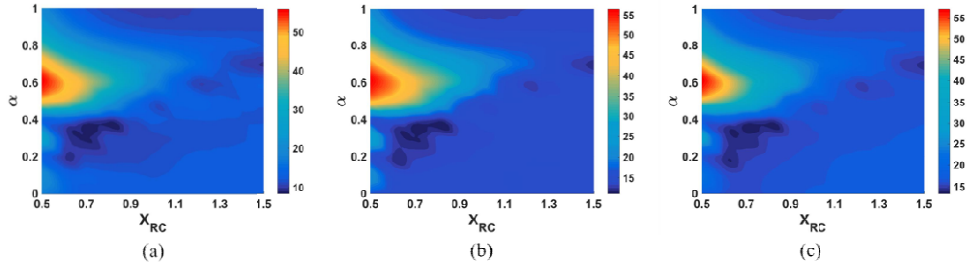

Fig. 4. EVM of a 30GBaud QPSK optical fiber communication system with a RC-FIR filter equalizing CD. (a)500 km SSMF. (b)1000 km SSMF. (c) 1500 km SSMF.

Similarly, the performance of a CD equalization optical fiber communication system can be optimized by the FWHM ($T$), i.e. $X_{GS}$, of Gaussian funcion $f_{GS}(t)$. Fig. 5 shows the EVM of a QPSK optical fiber transmission system with a GS-FIR filter. The EVM is least with $X_{GS} = 0.7$ even if symbol rate and fiber length vary.

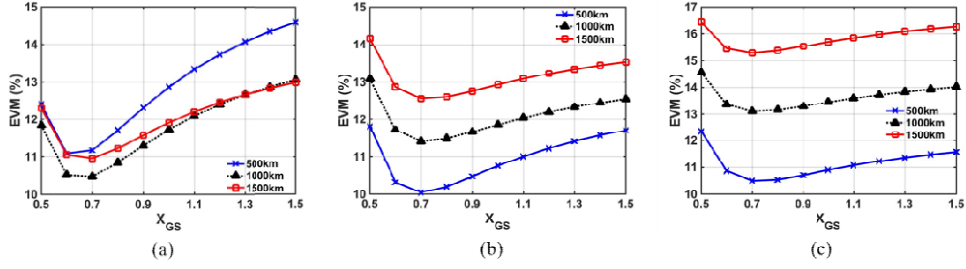

Fig. 5. EVM of QPSK signal with a GS-FIR filter equalizing CD.(a) 10GBaud QPSK. (b) 20GBaud QPSK. (c) 30GBaud QPSK.

## 4. Simulation results

For a S. Tsukamoto's transversal filter, the tap weights $a_k$ is derived by

$$a_k = h_{eq}(t)\big|_{t=t_k} = \text{ifft}\left\{H_{eq}(\omega)\text{Rect}(\omega)\right\}\big|_{t=t_k} \quad (9)$$

where $\text{ifft}\{\cdot\}$ is the inverted Fourier transform operation, $H_{eq}(\omega) = \exp\left(j\dfrac{D\lambda^2 L}{4\pi c}\omega^2\right)$, and $\text{Rect}(\cdot)$ is a rectangular function.

For a S. J. Savory's FIR filter, the tap weights $a_k$ is derived by

$$a_k = h_{eq}(t)\big|_{t=t_k} = \sqrt{\dfrac{jc}{D\lambda^2 L}}\exp\left(-j\dfrac{\pi c}{D\lambda^2 L}t^2\right)\text{Rect}(t)\big|_{t=t_k} \quad (10)$$

Fig. 6(a) compares the tap weights magnitude of transversal filter proosed by S. Tsukamoto, FIR filter, optimized RC-FIR filter and GS-FIR filter for a 20GBaud QPSK over 1000 km SSMF transmission system. Fig. 6(b) compares the frequency response of four filters. It can be seen from Fig. 6(b) that all the four filters can suppress the frequency aliasing for a single channel signal with limited bandwidth. The difference among four filters is obvious and will result in the difference of CD eqalization.

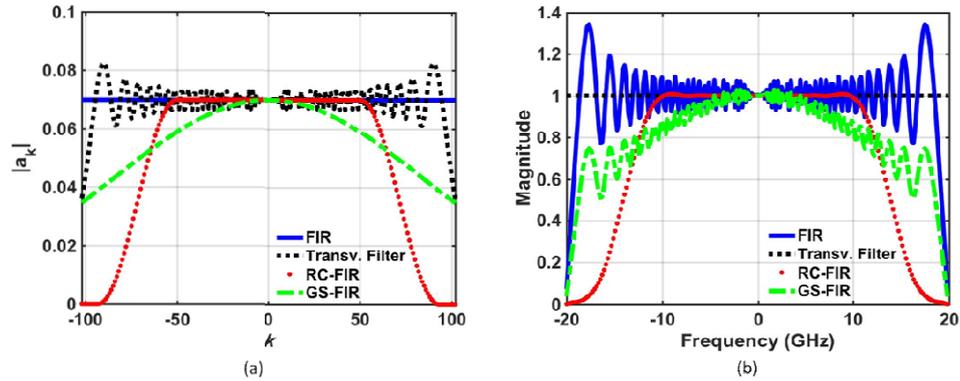

Fig. 6. (a)Tap weight magnitudes of FIR filter, transversal filter, RC-FIR filter and GS-FIR filter. (b) Frequency response of four filters.

The perormance of four filters, i.e. S. Tsukamoto's transversal filter, FIR filter, optimized RC-FIR filter and GS-FIR filter, for equlizing CD is compared based on VPI simultion. In the simulation, the setup of coherent optical fiber communication system is shown in Fig.1, and the fiber nonlinear effects and polarization mode dispersion are neglected since we mainly concentrate on the CD equalization in this work. 20GBaud QPSK, 16QAM and 32QAM transmission systems with different length of SSMF are investigated. EVM of the CD equalization signals employing the four filters are illustrated in Figs. 7-9 respectively. The results demonstrate that the CD equalization with both optimized RC-FIR filter and GS- FIR filter can improve the performance over Tsukamoto's transversal filter and FIR filter. The EVM of QPSK, 16QAM and 32QAM signals is lowed by 3~6 percent with the optimized RC-FIR filter for equalizing CD. The comparison of constellation diagrams is shown in Figs. 7-9 (b ~ e). The eye-diagrams of QPSK, 16QAM and 32QAM signals are clearest with the optimized RC-FIR filter. These results demonstrate that weighted FIR filter is an effective approach for improving the performance of CD equalization. The improvement can be sought by the optimization of a weighted function. Although we have investigated raised cosine function and Gaussian function as weighted functions for FIR filter in this paper, the weighted functions can be extended to other symmetric functions, such as super Gaussian function, hyperbolic secant function and etc.

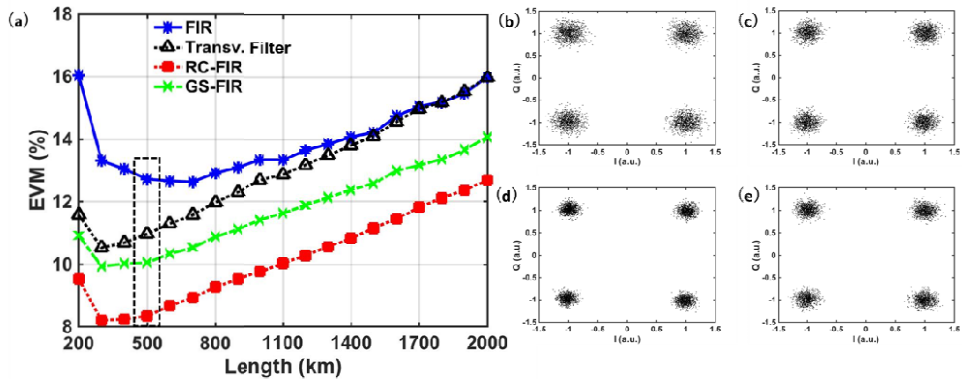

Fig. 7. Comparison of 40Gbit/s QPSK optical fiber communication system with four CD equalization filters. (a) EVM of QPSK signal after CD equalization. (b) Constellation diagram of QPSK signal after CD equalization with S. Tsukamoto's transversal filter. (c) Constellation diagram of QPSK signal after CD equalization with FIR filter. (d) Constellation diagram of QPSK signal after CD equalization with optimized RC-FIR filter. (e) Constellation diagram of QPSK signal after CD equalization with optimized GS-FIR filter.

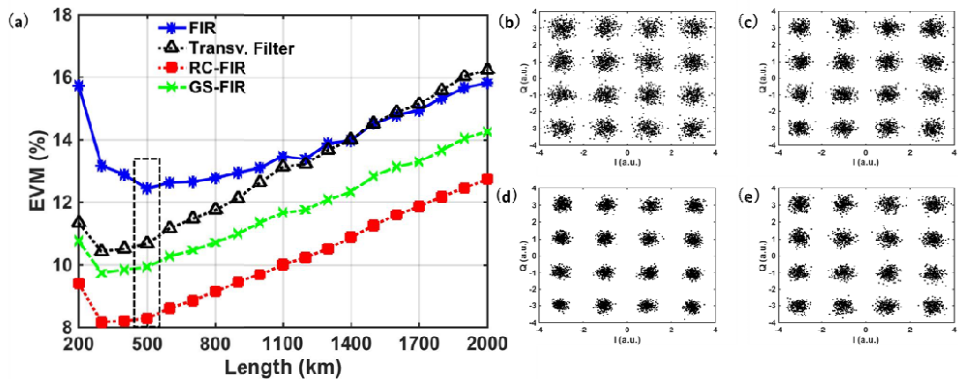

Fig. 8. Comparison of 80Gbit/s 16QAM optical fiber communication system with four CD equalization filters. (a) EVM of QPSK signal after CD equalization. (b) Constellation diagram of QPSK signal after CD equalization with S. Tsukamoto's transversal filter. (c) Constellation diagram of QPSK signal after CD equalization with FIR filter. (d) Constellation diagram of QPSK signal after CD equalization with optimized RC-FIR filter. (e) Constellation diagram of QPSK signal after CD equalization with optimized GS-FIR filter.

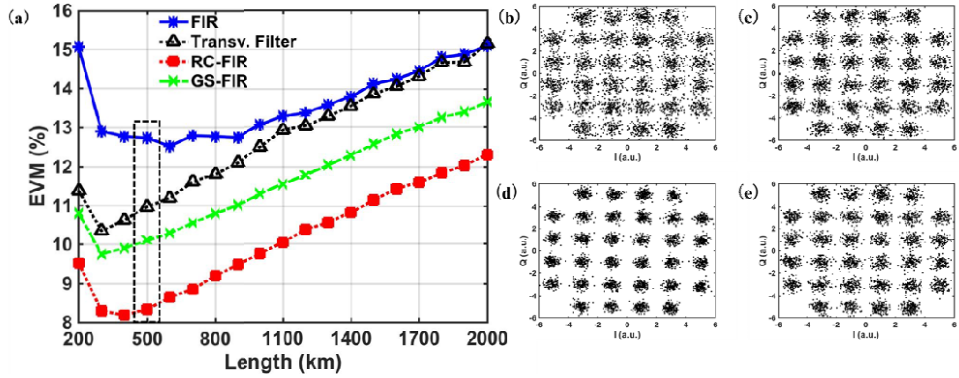

Fig. 9. Comparison of 100Gbit/s 32QAM optical fiber communication system with four CD equalization filters. (a) EVM of QPSK signal after CD equalization. (b) Constellation diagram of QPSK signal after CD equalization with S. Tsukamoto's transversal filter. (c) Constellation diagram of QPSK signal after CD equalization with FIR filter. (d) Constellation diagram of QPSK signal after CD equalization with optimized RC-FIR filter. (e) Constellation diagram of QPSK signal after CD equalization with optimized GS-FIR filter.

## 5. Conclusion

In this work, we propose weighted FIR filters for equalizing CD in time domain. As examples, a RC-FIR filter and a GS-FIR filter are investigated and optimized in term of the EVM of a CD equalization coherent optical fiber communication system. The optimized parameters of a RC-FIR filter and a GS-FIR filter are independent of modulation format, symbol rate and transmission distance. The simulation results of 20GBaud QPSK, 16QAM and 32QAM optical fiber communication system with the transmission length from 200 km to 2000km demonstrate that, both optimized RC-FIR filter and GS-FIR filter can improve the performance over previous reported FIR filter for CD equalization. The EVM of coherent communication system with optimized RC-FIR filter for CD equalization can be lowered at least 3 percent. The weighted FIR filter we propose in this work can also be extended to employ other symmetric weighted functions.

## Acknowledgements

This work is supported by National Natural Science Foundation of China (Grant No 61427813 and 61331010).